\documentclass[twocolumn,epjc3]{svjour3} 

\usepackage{color}
\usepackage{amssymb}
\usepackage{amsmath}
\usepackage{graphics}
\usepackage{epsfig}
\usepackage{bm}
\usepackage{xcolor}
\usepackage{slashed}
\usepackage[colorlinks,citecolor=blue,linkcolor=blue,urlcolor=blue]{hyperref}

\newcommand{\be}{\begin{equation}}
\newcommand{\bea}{\begin{eqnarray}}
\newcommand{\eea}{\end{eqnarray}}

\newcommand{\ee}{\end{equation}}

\begin{document}

\title{Aspects of the pseudo Chiral Magnetic Effect in 2D Weyl-Dirac Matter}

\author{
Ana Julia Mizher \thanksref{addr1,addr2}
        \and
        Sa\'ul Hern\'andez-Ortiz \thanksref{addr3}
        \and
        Alfredo Raya \thanksref{addr4}
        \and
        Cristi\'an~Villavicencio \thanksref{addr5}
} 

\institute{ Instituto de F\'isica Te\'orica, Universidade Estadual Paulista, 
R. Dr. Bento Teobaldo Ferraz, 271 - Bloco II, Barra-Funda
CEP: 01140-070, S\~ao Paulo, SP, Brazil 
            \label{addr1}
            \and
            KU Leuven Campus Kortrijk -- Kulak, Department of Physics, Etienne Sabbelaan 53 bus 7657, 8500 Kortrijk, Belgium 
            \label{addr2}
            \and
            Instituto de Ciencias Nucleares, Universidad Nacional Aut\'onoma de M\'exico, Apartado Postal 70-543, M\'exico Distrito Federal, C.P. 04510, Mexico    
            \label{addr3}
            \and
            Instituto de F\'isica y Matem\'aticas, Universidad Michoacana de San Nicol\'as de Hidalgo, Ediﬁcio C-3, Ciudad Universitaria, C.P. 58040, Morelia, Michoac\'an, Mexico 
            \label{addr4}
            \and
            Departamento  de  Ciencias  B\'asicas,  Facultad  de  Ciencias,
Universidad  del  B\'io-B\'io,  Casilla  447,  Chill\'an,  Chile. 
            \label{addr5}
}

\maketitle

\begin{abstract}
A connection is established between the continuum limit of the low-energy tight-binding description of graphene immersed in an in-plane magnetic field and the Chiral Magnetic Effect in Quantum Chromodynamics. A combination of mass gaps that explicitly breaks the equivalence of the Dirac cones, favoring an imbalance of pseudo-chiralities, is the essential ingredient to generate a non-dissipative electric current along the external field. Currents, number densities and condensates generated from this setup are investigated for different hierarchies of the energy scales involved.
\end{abstract}

\section{Introduction}
\label{intro}

The chiral magnetic effect (CME), predicted in Ref.~\cite{Kharzeev:2007jp,Fukushima:2008xe} to occur in the quark gluon plasma produced in heavy ion collisions, has a deep connection with the vacuum structure of quantum chromodynamics (QCD), its topology and symmetries. The non-dissipative current produced by this mechanism points out in the direction of the magnetic field generated in non-central collisions and is a direct consequence of a chiral imbalance. Such imbalance can only occur in certain domains where the gauge field configurations are topologically non-trivial. Interactions of fermions with these fields result in a chirality flip.

 The difficulty to extract information about the early stage in HIC -- due to unknown electromagnetic properties of the medium, the out-off-equilibrium regime and the lack of effective transport descriptions -- makes the CME so far an unique attempt to trace a mechanism that connects some peculiar quantum properties of QCD to a macroscopic observable.
Besides contributing for a more complete picture of QCD, its confirmation would have implications in our understanding about the early universe and baryogenesis.
Because of that, it has received considerable attention from the theoretical, experimental and lattice communities in high energy physics but in spite of all efforts, a conclusive observation of the mechanism is still missing. In effect, results from the CMS collaboration~\cite{Khachatryan:2016got} comparing angular correlation between Pb-Pb collisions and p-Pb collisions have challenged previous results from the STAR collaboration~\cite{Abelev:2009ac} that seemed to have observed the CME. 
In order to clarify this issue, new technology on background analysis is being carried out \cite{Sirunyan:2017quh} and new observables have been proposed \cite{Sirunyan:2017quh,Magdy:2017yje}. Remarkably, the 2018 run of RHIC includes isobar collisions in order to disentangle the chiral magnetic effect from background sources \cite{Skokov:2016yrj}.

In a different order of ideas, the fast-growing family of Weyl-Dirac materials that have been discovered in the last few years have allowed to test an analog of this mechanism in a condensed matter environment~\cite{Li:2014bha}. 
In this kind of materials, the complex interaction between the charge carriers and the background lattice can be effectively represented considering the former as quasiparticles that obey relativistic-like equations of motion, with the velocity of light replaced by the corresponding Fermi velocity. 
In this way, it is possible to define chirality for these electrons and to construct the analogy with QCD~\cite{Gusynin:2007ix,Vozmediano:2010zz,Cortijo:2011aa}. 

In search for inducing a chiral splitting of charge carriers in Weyl-Dirac materials, an experiment was proposed in Ref.~\cite{Li:2014bha} such that by applying a parallel electric and  magnetic field  to a (3+1)D sample of zirconium pentatelluride (${\rm ZrTe}_5$), the observation of a negative magnetoresistence signals the presence of a chiral anomaly \cite{Son:2012bg} and the generation of a non-dissipative electric current. This represents yet another novel avenue allowing for interdisciplinary investigation connecting condensed matter and high energy physics.
Nevertheless, the Weyl-Dirac behavior of the charge carriers in ${\rm ZrTe}_5$ is still controversial. While the authors of Ref.~\cite{Li:2014bha} established the {\em ultrarelativistic} behavior of charge carriers through  angle-resolved photoemission spectroscopy (ARPES), further investigation has opened the possibility that this  may not be actually the case after all~\cite{PhysRevB.95.195119}.
A more detailed understanding of the mechanism in the context of condensed matter is still needed and realizations of CME in different materials are welcome. 
 In effect, it has been posteriorly observed in other materials \cite{CME_DiracWeyl} and alternative mechanisms have been proposed in order to generate the CME in $3+1$D Weyl semimetals \cite{Cortijo:2016wnf} but no experimental realization have been achieved so far.

In this paper, we extend the work presented in Refs.~\cite{Mizher:2013kza,Mizher:2016mfq}, where some of us proposed an electromagnetic analogue of the CME in Weyl-Dirac materials in two spatial dimensions motivated by graphene and referred to as the {\em pseudo-chiral magnetic effect} (PCME). 

It is natural to search for this type of analogy because  quantum electrodynamics in (2+1)D, dubbed as QED$_3$, has been widely used as a toy model for QCD inasmuch as it describes confinement and chiral symmetry breaking~\cite{PhysRevLett.57.957,Appelquist,Fischer,BashirRaya,PhysRevD.94.125009}. 
This is due to the fact that at high temperature, any field theory can be dimensionally reduced and, on the other hand, a non-Abelian three dimensional gauge field theory \emph{abelianizes} for a large number of flavors~\cite{Pisarski:1984dj}. 
With the physical realization of graphene and other planar materials containing Dirac electrons, QED$_3$ was promoted from being a toy model to really describe physical systems and analogues of high energy physics constructed on table-top experiments started to take place. 

To describe the PCME, in Refs.~\cite{Mizher:2013kza,Mizher:2016mfq} we constructed a Lagrangian for relativistic fermions in (2+1)D including some effective interactions that simulate a chiral imbalance. The result is an electric current generated in the direction of an external magnetic field applied in-plane. 
Here we extend this study investigating other relevant quantities: the number densities and condensates that we found to be present and the current  in certain limiting cases of magnitude of the magnetic field as compared with temperature, mass gaps and chemical potentials. 

The paper is organized as follows: in Section \ref{model} we give a brief review of how Dirac fermion emerge from graphene and in this context we present our Lagrangian and describe our model. 
In Section~\ref{propagator} we discuss the propagator of a fermion in the presence of a magnetic field based on the Schwinger proper time method. 
In Section~\ref{currents} we present our results on the currents, densities and condensates. In Section~\ref{sec:mu5} we explain our model in terms of the Fermi liquid model, commonly applied in condensed matter physics. 
In Section~\ref{conclusions} we make our final remarks. We complement the article with three appendixes.

\section{The effective model}
\label{model}

The essence of the CME lies on the fact that the vacuum of the gauge sector in QCD is actually a superposition of vacua including configurations with non-trivial topology, and the interaction of these gauge fields with quarks can cause a flip on their chirality. In that case, an imbalance between chiralities occurs and, inside a certain domain where the topological gauge fields act, one has a difference in the number of left- and right-quarks, whose abundance and sign depends on the details of the topology of the gauge fields. The large magnetic field present in non-central heavy-ion collisions triggers the CME by generating
an electric current along its  direction.
Therefore, the chiral imbalance and the magnetic field constitute the two pillars of the CME.

Planar graphene-like materials  provide elements we need for a robust analogy. This kind of materials have the remarkable property that the charge carriers behave as Dirac electrons, which means that they have a linear dispersion relation. 
Besides, the continuum limit of its tight-binding Hamiltonian (or Lagrangian) 
 exhibit Time- and Space-inversion symmetries such that massless charge carriers are, in fact, chiral. 
Theoretically, it is possible to break these symmetries if we consider special currents in the Lagrangian, as will be clear in a moment. 

\begin{figure}[t]
\centering
\includegraphics[width=0.45\textwidth]{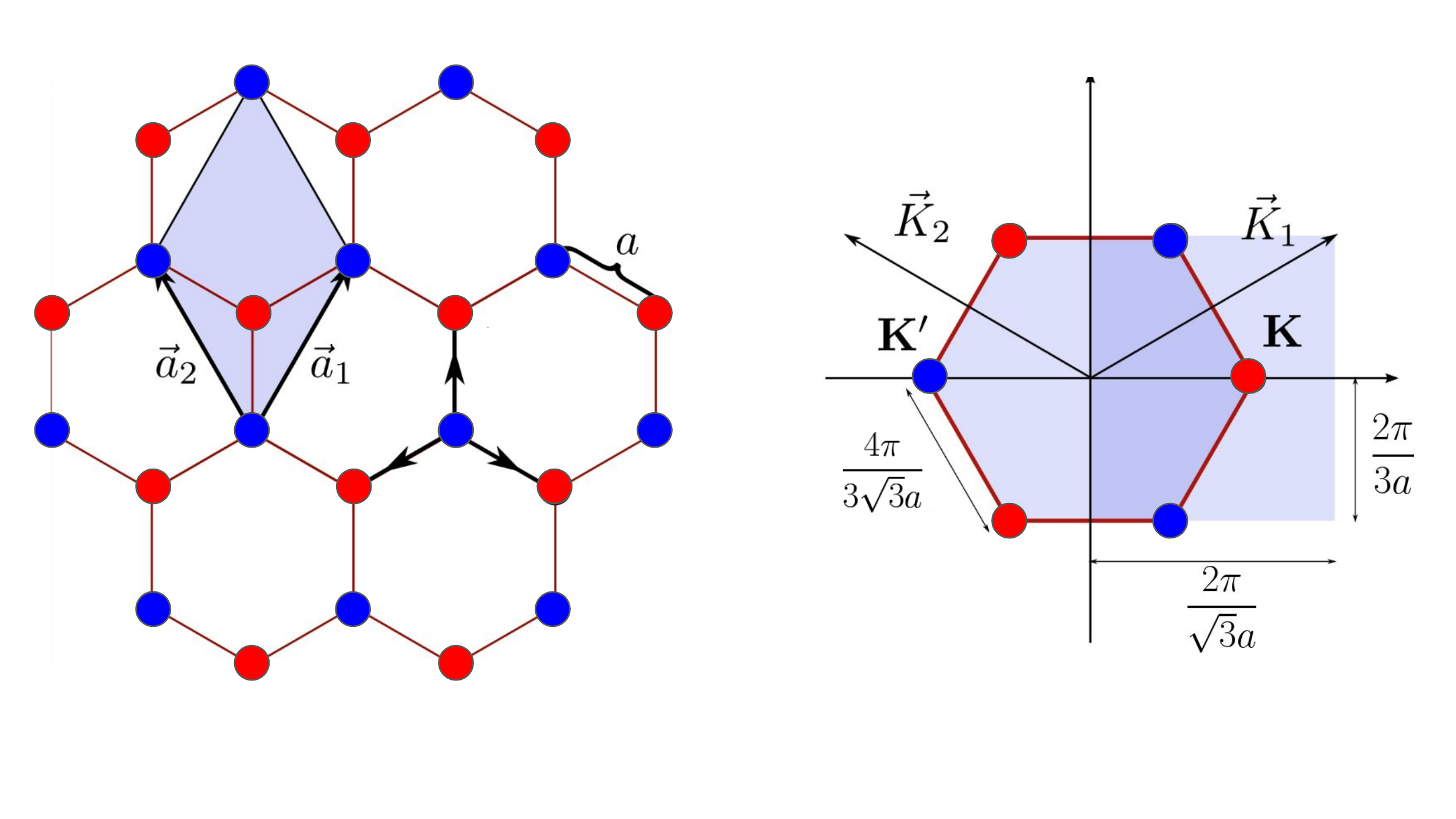}
\caption{Crystal structure of the honeycomb array of Weyl-Dirac materials in real (left panel) and reciprocal (right panel) spaces.
Primitive vectors are explained in the text. The interatomic distance is $a$}
\label{red}
\end{figure}

Let us briefly describe the low-energy continuum dynamics of graphene-like materials. 
For a more detailed description one counts on several books and reviews on the subject, including Refs.~\cite{Gusynin:2007ix,Vozmediano:2010zz,Cortijo:2011aa,Katsnelson} and references therein.
The honeycomb array of these materials does not have the structure of a Bravais lattice required for a tight-binding description. Nevertheless, it can be regarded in terms of two equivalent triangular sub-lattices $A$ and $B$ which are Bravais lattices, in which the nearest neighbor of a carbon atom belonging to sub-lattice $A$ are all atoms in the sub-lattice $B$ and vice versa, as shown in Fig.~\ref{red}.  
With the primitive vectors  ${\mathbf a}_1=a(1/2,\sqrt{3}/2)$, and ${\mathbf a}_2=a(-1/2,\sqrt{3}/2)$, where $a$ denotes the interatomic distance, and correspondingly in reciprocal space ${\mathbf K}_1=2\pi/a(1,\sqrt{3}/3)$ and ${\mathbf K}_2=2\pi/a(-1,\sqrt{3}/3)$, from Bloch theorem it is straightforward to find that the energy-momentum dispersion relation is

\begin{figure}[t]
\centering
\includegraphics[width=0.4\textwidth]{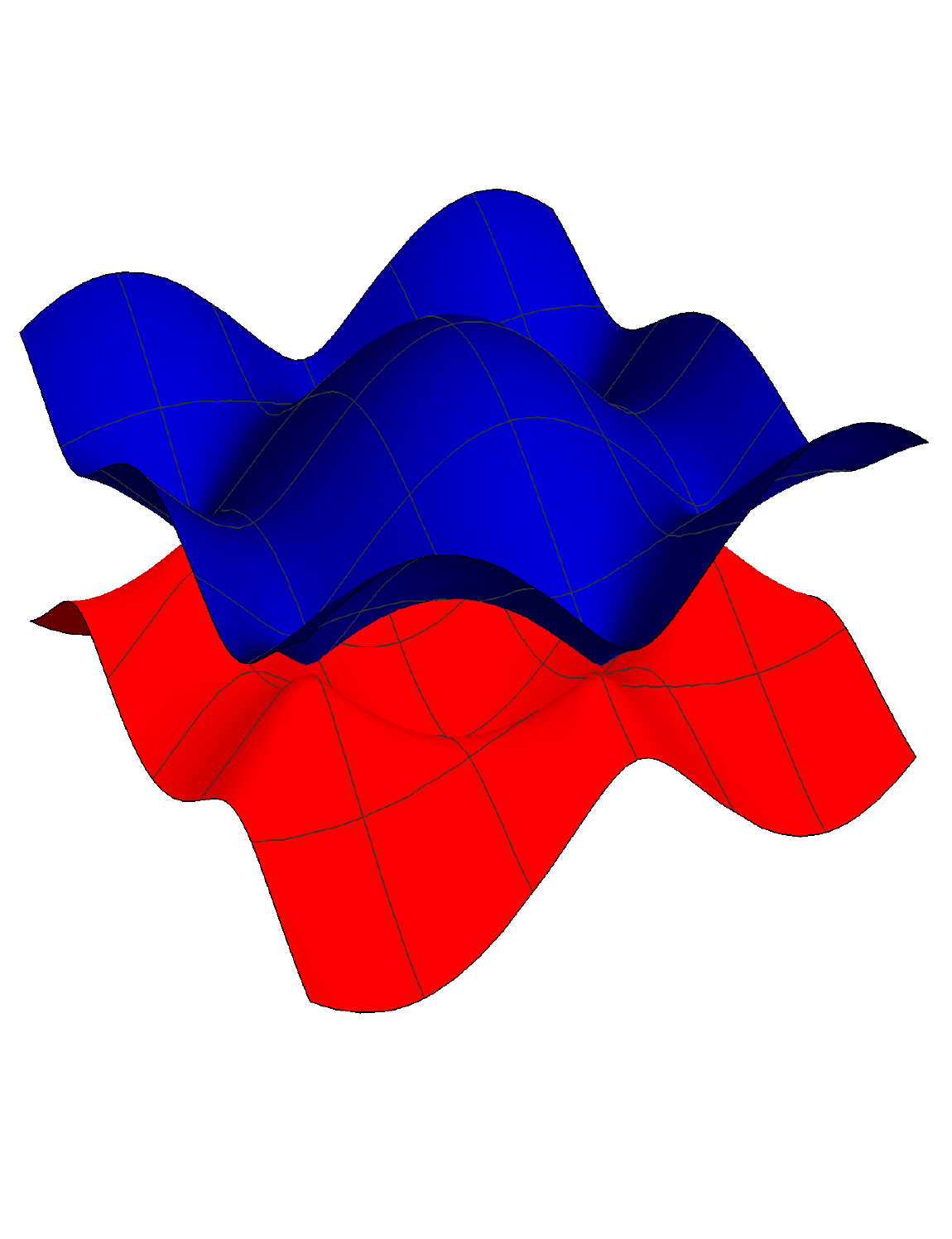}
\caption{Energy-momentum dispersion relation for the honeycomb array.
There are six high-symmetric points in the Brillouin zone where valence and conducting bands touch.
Only two of them, the Dirac points $K$ and $K'$, are inequivalent. }
\label{conos}
\end{figure}

\begin{multline}
 E(k_x,k_y)= 
 \pm t\Bigg[3+2\cos\left({\sqrt{3}k_y a}\right)\\
 +4\cos\left({\frac{\sqrt{3}}{2}k_y a}\right)\cos\left({\frac{3}{2}k_x a}\right)\Bigg]^{1/2}
\end{multline}
where $t$ is a constant that encodes the magnitude of the interaction (the hopping parameter) and $k_i$ correspond to the momentum components in the two spatial directions. This function is plotted in Fig.~\ref{conos}. One can notice that the valence and conduction band touch in six points, generating no gap. Those are high symmetry points in the Brillouin zone of which only two are inequivalent, the so-called Dirac points $K$ and $K'$.
Expanding the energy around these points, one can verify that the dispersion relation is linear, $E = \hbar v_F k$, and we can think of charge carriers as massless Dirac electrons, with the speed of light replaced by the Fermi velocity. Therefore, the tight-binding  description of monolayer graphene corresponds in the continuum to a massless version of the fermion sector of QED$_3$,  but photons are allowed to move in (3+1) dimensions. 
 The influence of external electromagnetic fields as usual is added via minimal coupling.s

The symmetry between the two sublattices allows to represent the charge carriers as four component spinors, with two indices corresponding to the sub-lattice {\em flavor} or pseudo-spin and the Dirac point (valley) index, respectively. 
Working in Weyl representation, the gamma matrices can be explicitly written:
\begin{eqnarray} \nonumber
&&\gamma^0=\left(
\begin{array}{cc}
0 & I_2 \\
I_2 & 0
\end{array}
\right),\quad \gamma^i=\left(
\begin{array}{cc}
0 & \sigma_i \\
-\sigma_i & 0 
\end{array}\right) , \qquad i=1,2,3\\[10pt]
&&\gamma^5=\left(
\begin{array}{cc}
-I_2 & 0 \\
0 & I_2
\end{array}
\right).
\end{eqnarray}
Notice that in this case, the free massless Dirac Lagrangian 
\begin{equation}
\mathcal{L}=\bar{\psi}\ i{\not\! \partial}\ \psi \; ,\label{massless}
\end{equation}
with ${\not \! \partial}=i\gamma^\mu\partial_\mu$ and $\mu=0,1,2$, is invariant under the chiral-like transformations
\begin{equation}
\qquad \psi \to e^{i\theta \gamma^5}\psi \;, \qquad \psi \to e^{i\theta' \gamma^3\gamma^5}\psi \,.
\label{eq:chiral}
\end{equation}
Preserving the symmetries in Eq.~(\ref{eq:chiral}), a couple of terms, bilinear on the fermion fields, can be added, $m_o\bar\psi \gamma^3\gamma^5 \psi$ 
and $m_e \bar{\psi} \gamma^3\psi$. The first term is known as Haldane mass and in QED$_3$ it is directly related to a Chern-Simons (CS) effective action upon integrating fermionic degrees of freedom. 
The second, proportional to $m_e$, corresponds to sublattice symmetry breaking.

Let us now discuss the role of the parity anomaly and how it connects to the aforementioned masses. The ABJ anomaly in 3+1D can be interpreted as production of Weyl fermions in the presence of external electric and magnetic fields \cite{Ambjorn:1983hp}. This was initially proposed in a nuclear physics context and in \cite{Nielsen:1983rb} this interpretation was extended to condensed matter systems. The authors show that the contribution to the anomalous electric current coming from fermions with different chiralities have opposite sign and, since it was previously shown that in any lattice theory with locality, chiral fermions appear in pairs of opposite chirality \cite{Nielsen:1981hk} - fermion doubling - the net current vanishes.

In odd dimensions, it is known that the conservation laws for fermionic currents do not present the chiral anomaly, but it might still emerge abnormal currents for each specie of fermions due to the parity anomaly \cite{Niemi:1983rq}. In \cite{Semenoff:1984dq} the author proposes a 2+1D analogue of \cite{Nielsen:1983rb} in a graphene-like system. Introducing a gap between the bands induced by a sublattice symmetry breaking, the author show that the contribution from the two inequivalent Dirac points cancel out, similarly to what happens in 3+1D. The two Dirac points are therefore the analogues for chirality.

In our work we propose something similar, but we look for a Lagrangian structure where the masses for the two valleys are different and therefore for suitable values of the parameters the net current does not vanish. We obtain this by the combination of Haldane mass $m_o$ and the mass $m_e$, which carries an identical structure as the one Semenoff uses in \cite{Semenoff:1984dq}. The anomaly is therefore parametrized in $m_\pm$ in the same way that the chiral anomaly is parametrized in $\mu_5$ in the original chiral magnetic effect. The underlying Hamiltonian is therefore, equivalent to the one in \cite{Semenoff:1984dq} but with different masses for each Dirac point. Further, as we explain below, we promote the interaction of this system with an external magnetic field via minimal coupling and consider thermal effects.

To complete our analogy, we consider an external magnetic field aligned along the graphene membrane described by the vector potential $A_3^{\mathrm{ext}}=By$, where $y$ represents the second space coordinate along the graphene plane and we assume $B>0$. This field is assumed to be classical and does not play a role in quantum corrections. Since it is external, in our setup $A_\mu$ lives in a bulk rather than in a two-dimensions sheet. Other field theory approaches for QED in planar condensed matter systems have been presented previously considering an explicit treatment of the gauge sector, see for instance the Pseudo Quantum Electrodynamics (PQED) \cite{Marino:1992xi,Alves:2013bna,Marino:2014oba,Nascimento:2015ola} and the Reduced Quantum Electrodynamics (RQED) \cite{Gorbar:2001qt}.  
In the present work we do not deal with the Maxwell term and do not consider quantum corrections coming from the gauge sector and therefore we do not make use of PQED/RQED. However, in a work in progress we are currently investigating if within this more complete framework $m_o$ can be dynamically generated by the presence of parallel electric and magnetic fields. 
One of us have checked some formal aspects of the theory in mixed dimensions \cite{Dudal:2018mms,Dudal:2018pta} in order to prepare for the numerical calculations and we expect to report soon.
In our approach we just consider the fermion sector and the interaction with the classical field. We restrict the dynamical term $\partial \psi /\partial x_3 =0$ but we preserve the fermion interaction with $A_3$.

\begin{figure}[t]
\centering
\includegraphics[width=0.45\textwidth]{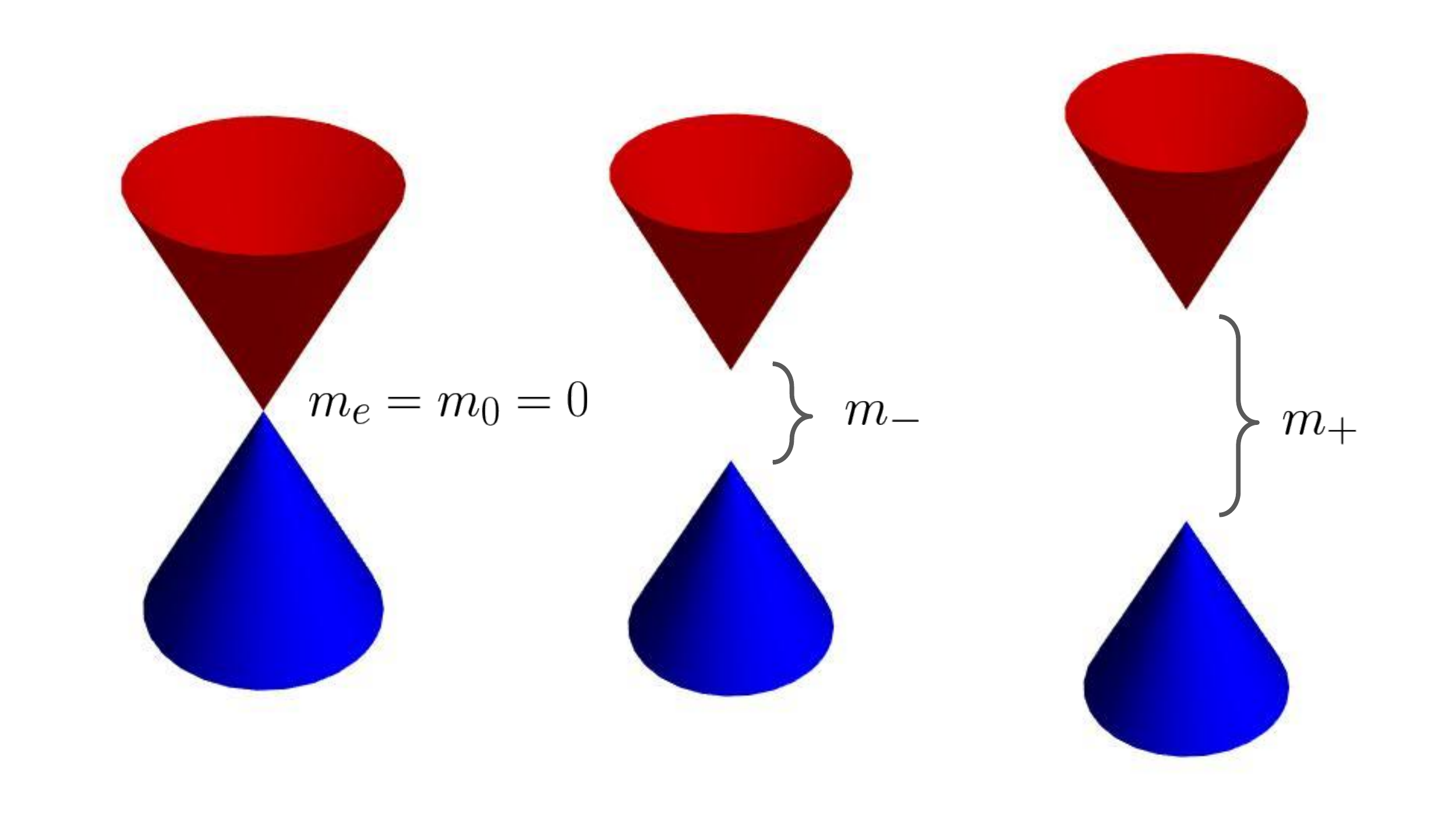}
\caption{Sketch of masses for the fermion fields in Lagrangian~(\ref{lagrangiano}). One specie becomes heavier and the other one lighter as $m_0$ grows.}
\label{Gaps}
\end{figure}

The scenario described above can be represented by the Lagrangian:
\begin{equation}
\mathcal{L_F}=\bar{\psi}\left[i\slashed{D}+\left(eA_3^{\mathrm{ext}}-m_e\right)\gamma^3-m_0\gamma^3\gamma^5\right]\psi \; ,\label{original}
\end{equation}
where ${\slashed{D}}=\left(\partial_0-i\mu,v_F\bm{\nabla}\right)$, $e=-|e|$ is the fundamental charge, $v_F$ is the Fermi velocity of quasiparticles in the crystal which from now onward we set in the natural units of the system (namely, $v_F=1$), and $\mu$ the chemical potential. In the Weyl representation for the gamma matrices, the matter content of the theory can be unveiled by introducing the chiral-like projection operators $\chi_\pm=\frac{1}{2}\left(1\pm\gamma_5\right)$ which verify $\chi_{+}+\chi_-=1,$ and $\chi_\pm^2=\chi_\pm,$ and thus allow us to write 
\begin{equation}
\mathcal{L_F}=\sum_{\chi=\pm}\bar{\psi}_\chi\left[i\slashed{D}+\mu\gamma^0+\left(eA_3^{\mathrm{ext}}-m_\chi\right)\gamma^3\right]\psi_\chi \; ,\label{lagrangiano}
\end{equation}
where $m_\pm=m_e\pm m_0$ and the ``left-'' and ``right-handed'' fields are $\psi_\pm=\chi_\pm \psi$. Hence, the Lagrangian represents a system where the two different charge carrier species become non-degenerate in gaps: one of them becomes lighter and the other heavier as $m_o$ grows (see Fig.~\ref{Gaps}).

\section{Propagator}
\label{propagator}

In order to have access to currents, densities and condensates arising in our system, we require the corresponding Green function for fermions from the Lagrangian in Eq.~(\ref{original}), which already takes into account the influence of the in-plane
external magnetic field, the chemical potential and temperature. We start by splitting
the complete Green function into separate Green functions 
for each chirality, $G_\pm$,  such that 
\begin{equation} 
    G\left(r,r^\prime\right)=\frac{1+\gamma^5}{2}G_+\left(r,r^\prime\right)+\frac{1-\gamma^5}{2}G_-\left(r,r^\prime\right)\; .\label{full}
\end{equation}
Since we are interested in exploring this effect at room temperature we must include thermal effects. This is introduced in the usual way by replacing the zeroth component of the momentum with the Matsubara frequencies  $k_0 \rightarrow i\omega_n = i(2n+1)\pi T$ and momentum integrals are replaced by sums over these frequencies, $\int dk_0 f(k_0) \rightarrow i2\pi T \sum_n f(i\omega_n)$. 
Secondly the inclusion of the external magnetic field can be treated within the Schwinger proper time method~\cite{PhysRev.82.664}. 
Recall, however, that in the presence of  finite chemical potential, some care must be taken in the range of integration in the proper-time integrals in order to guarantee their correct convergence~\cite{PhysRevD.42.2881}. 
This can be done by the mnemonic rule
\begin{equation}
    \int_0^\infty ds\ g(s)\rightarrow \int_{-\infty}^\infty ds\ r_s\,g(s) \;,
\end{equation}
where the regulator reads
\begin{equation}
    r_s =\text{sign}(s)\,\theta(s\,\omega_n\mu).
\end{equation}
Thus, we express the resulting propagator for each  chirality as \cite{Mizher:2013kza}:
\begin{equation}
    G_\pm\left(r,r^\prime\right)=iT\sum_n\int\frac{d^2k}{\left(2\pi\right)^2}e^{-ik\cdot\left(r-r^\prime\right)}\tilde{G}_n\left(k;M_\pm\right)\; ,\label{chiral}
\end{equation}
with
\begin{eqnarray}\label{eq:propagator}
    \tilde{G}_n\left(k;M_\pm\right) & = &  \int_{-\infty}^\infty ds\ r_s \,e^{isK_\parallel^2-i\left[k^2+M_\pm^2\right] \tan(eBs)/eB} \nonumber \\
     &&\times \bigg\{ \slashed{K}_\parallel \left[1+\gamma^2\gamma^3\tan(eBs)\right] \nonumber \\
     & & \qquad - \left[k_2\gamma^2+M_\pm\gamma^3\right]\sec^2(eBs)\bigg\}\; ,
\end{eqnarray}
where we made use of the shorthand notation $M_\pm=\frac{1}{2}\left(y+y'\right)eB+m_\pm$ with the notation $(r^1,r^2,r^3)\equiv (x,y,z)$. We also define $K_\parallel=\left(i\omega_n+\mu,k^1,0\right)$ which is the parallel component to the magnetic field, oriented in the $x$-direction. 
Notice that the Green function is non-local due to the term $M_\pm$.

In the particular case where $\mu<\pi T$, the proper-time integral can be Wick-rotated in a simple way (see \cite{Mizher:2013kza} for more details). 
In this case, the propagator in Eq. (\ref{eq:propagator}) can be written as
\begin{eqnarray}\label{eq:propagator2}
    \tilde{G}_n\left(k;M_\pm\right) & = &  -i\int_{0}^\infty ds\,e^{-sK_\parallel^2-\left[k^2+M_\pm^2\right] \tanh(eBs)/eB} \nonumber \\
     &&\times \bigg\{ \slashed{K}_\parallel \left[1-i\gamma^2\gamma^3\tanh(eBs)\right] \nonumber \\
     & & \qquad - \left[k_2\gamma^2+M_\pm\gamma^3\right]\text{sech}  ^2(eBs)\bigg\}.
\end{eqnarray}
So, for now onward we consider this regime, where the system is diluted enough to fulfill the condition $\mu<\pi T$.
Below we derive expressions for currents, densities and condensates accessible  from this Green function.

\section{Currents, densities and condensates}
\label{currents}

From the full Green function in Eqs.~(\ref{full}), (\ref{chiral}) and (\ref{eq:propagator}), we are in position of obtaining several physical observables as functions of the magnetic field $B$, temperature $T$, chemical potential $\mu$ as well as the charge carrier species gaps $m_\pm$ with different hierarchies among these scales. 
The non-local nature of the Green function in Eq. (\ref{eq:propagator}) indicates that finite size effects are important to be considered. 
The relevant expectation value of the objects 
\begin{equation}
J_\Gamma(y) = \langle\bar\psi\Gamma\psi\rangle = -\text{tr}[\Gamma G(r,r)]\;,
\end{equation}
can be defined with respect to a given origin, which we choose, without loss of generality, to be the center of the layer in the $y$-direction. The setup is symmetric with respect to the $x$-direction.
The appropriate value can be considered then as the average in space, namely
\begin{equation}\label{Jpromedio}
\bar J_\Gamma = \frac{1}{L_y}\int_{-L_y/2}^{L_y/2}dy \;J_\Gamma(y).
\end{equation}
Let us comment about the apparent dependence of the current on the coordinate $y$. A translation in our propagator can be written as $A_3=By \rightarrow B(y-y_0)$. Note that this is in fact equivalent to shift the mass $m_e \rightarrow m_e+eBy_0$ and therefore different integration limits yield to the same result as far as $m_e$ is correctly shifted. In the absence of magnetic field this parameter corresponds to the gap generated by sublattice symmetry breaking obtained for example placing graphene over a boron nitride substrate. If a magnetic field is turned on, $m_e$ plays the role of a bare mass that can absorb an eventual constant added to the gauge field and its physical value must be fixed by some observable like the electric current. 

This is similar for instance to the procedure adopted in \cite{electromagnetic}, where a constant uniform electric field generated from a static potential produces a non-local term, which is considered as a coordinate-dependent chemical potential and must be fixed by a suitable quantity like the number density.

All the relevant quantities can be expressed in a closed analytical form, instead of integrals in proper-time. 
We explore two regimes for the magnetic field with respect to the temperature and chemical potential. More specifically, the 
relevant scale to which we compare the field strength is $(\pi T)^2-\mu^2$, namely, the quantity that needs to be positive to enable for a Wick rotation. 
Details of the asymptotic low- and high-magnetic field approximations are described in the appendices.
The case of pure magnetic contribution for the condensates is be treated separately.

\subsection{Currents}

First we obtain the electric currents for each sub-lattice and the chiral current according to
\begin{eqnarray}
    j_i(y) &=& e \, \langle \bar{\psi} \gamma_i \psi  \rangle
            =- e \,{\rm tr}[\gamma_i G(r,r)]\;,\\
    j_{i 5}(y) &=& e\, \langle \bar{\psi} \gamma_i \gamma_5 \psi  \rangle 
            = -e\, {\rm tr}[\gamma_i \gamma_5 G(r,r)]\;,
\end{eqnarray}
with $i=1,2$.
Tracing over gamma matrices, one can notice that the only non-vanishing components of the currents are $j_1$ and $j_{15}$ along the magnetic field direction. This is the essence of the PCME. 
We express these currents  as
\begin{eqnarray}
    j_1(y) &=& j(y-y_+)-j(y-y_-)\;,\\
    j_{15}(y) &=& j(y-y_+)+j(y-y_-)\;,
\end{eqnarray}
with $y_\pm=m_\pm/eB$ and the function $j(\eta)$ defined as
\begin{eqnarray}\label{corriente}
    j(\eta)&=&-i\frac{e^2B}{\pi}T \sum_n \int_{-\infty}^\infty ds\,
    \left(\omega_n-i\mu\right) 
		\left[\frac{\tanh(eBs)}{eBs}\right]^\frac{1}{2}		\nonumber\\
		&& \times\mathrm{exp}\left(-\left[s\left(\omega_n-i\mu\right)^2+eB\tanh(eBs)\eta^2\right]\right)\,. \nonumber\\
\end{eqnarray}
The Matsubara sums can be performed straightforwardly using the definition of the Jacobi $\Theta_3$-function and then, using the inversion relation for the Jacobi function (see \ref{formulas}) we can write
\begin{eqnarray}\label{jnum}
    j(\eta) &=&-\frac{e^2B}{4\pi^{3/2}}\int_0^\infty \frac{ds}{s^{3/2}}\frac{\mathrm{exp}\left[-eB\tanh(eBs)\eta^2\right]}
    {\sqrt{eBs/\tanh(eBs)}} 
    \nonumber \\ &&
    \times \frac{\partial}{\partial \mu}
    \Theta_3\left(\frac{1}{2}-\frac{i\mu}{2\pi T},\,\frac{i}{4\pi s T^2}\right).
\end{eqnarray}

\begin{figure}[t]
\centering
\includegraphics[width=0.4\textwidth]{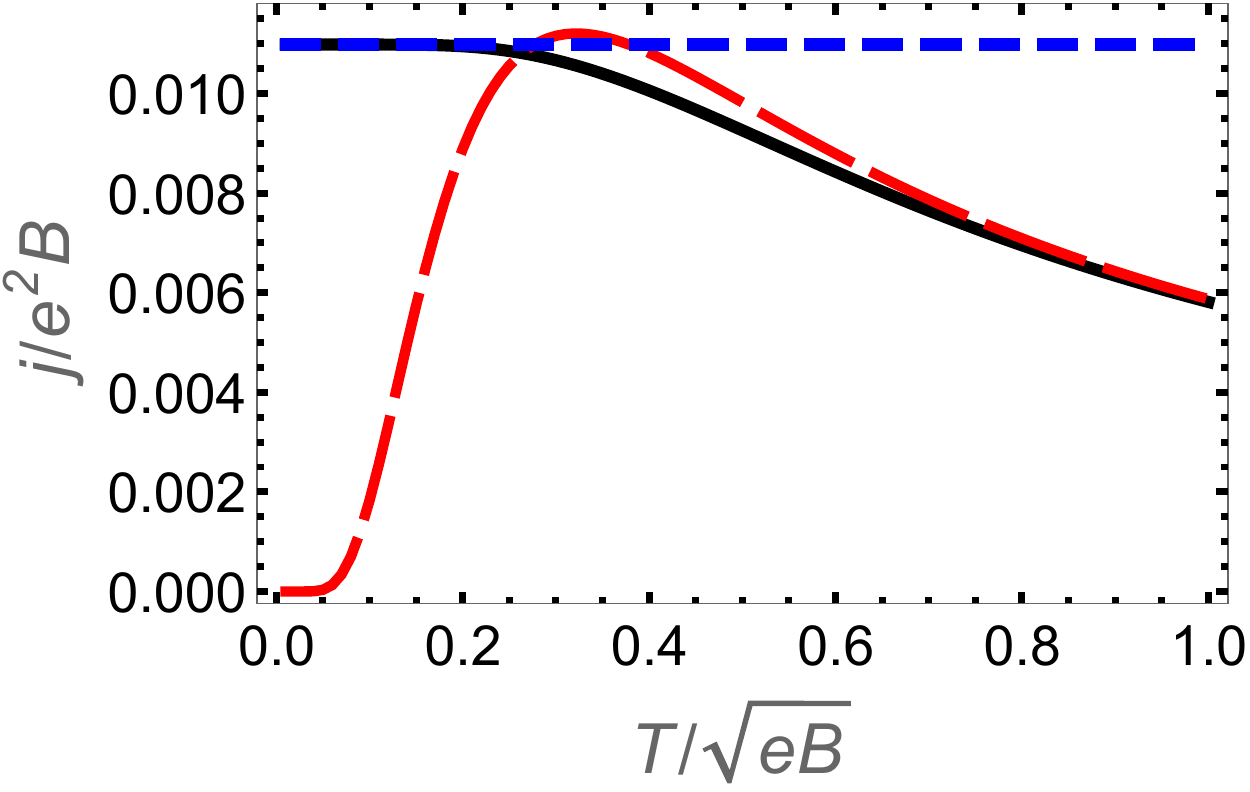}
\caption{Behavior of $j(\eta)$ as a function of $T/\sqrt{|eB|}$ for  $\sqrt{|eB|}\eta = 0.5$ and $\mu/\sqrt{|eB|}=\pi/40$. 
The solid (black) curve corresponds to the numeric expression in Eq.~(\ref{corriente}).
The long-dashed (red) curve corresponds to the weak field approximation in Eq.~(\ref{jweak}). 
The short-dashed (blue) curve corresponds to the strong field approximation in Eq.~(\ref{jstrong}). }   
\label{fig1}
\end{figure}

\bigskip
The approximation of low and high magnetic field is obtained from Eq. (\ref{corriente}). 
For $|eB|\ll(\pi T)^2-\mu^2$, 
\begin{eqnarray}\label{jweak}
    j(\eta) = \frac{e^2B}{2\pi}\left[n_F\left(eB\,\eta-\mu\right)-n_F\left(eB\, \eta+\mu\right)\right],
\end{eqnarray}
where $n_F(x)=(1+e^{x/T})^{-1}$ is the Fermi-Dirac distribution function.
For $|eB|\gg (\pi T)^2-\mu^2$, 
\begin{equation}\label{jstrong}
    j(\eta)=\frac{e^2B}{\sqrt{|eB|}}\, \frac{\mu}{\pi^{3/2}} e^{-|eB|\eta^2}\;.
\end{equation}
A comparison of the weak and strong field expansion and the full numerical result with all the parameters scaled with $|eB|$ is shown in Fig.~\ref{fig1}. 
We can observe that the approximations are in good accordance with the numerical expression.
It is interesting that in fact for strong magnetic field (low temperature) the dependence of the currents on the temperature is negligible.

\subsection{Number  densities}

We proceed the same as in the case of electric and chiral currents.
The  number density and the chiral number density are defined as
\begin{align}
n = & \left \langle \psi^\dagger \psi \right \rangle
        = -{\rm tr}[\gamma_0 G(r,r)], \\
n_5 = &\left \langle \psi^\dagger\gamma_5 \psi \right \rangle
        = -{\rm tr}[\gamma_0\gamma_5 G(r,r)].
\end{align}
and we can express these quantities as
 \begin{eqnarray}
    n(y) &=& \nu(y-y_+)+\nu(y-y_-)\;,\label{eq:nu}\\
    n_{5}(y) &=& \nu(y-y_+)-\nu(y-y_-)\;,
\end{eqnarray}
where the auxiliary function $\nu\left(\eta\right)$ is defined as
\begin{eqnarray}\label{numero}
    \nu(\eta)&=&
    -i\frac{1}{\pi}T \sum_n \int_{0}^\infty \frac{ds}{s}
    \left(\omega_n-i\mu\right)
    \left[\frac{eB s}{\tanh(eBs)}\right]^\frac{1}{2}\nonumber\\
&& \hspace{-5mm}\times   
        \mathrm{exp}\left(
        -\left[s\left(\omega_n-i\mu\right)^2+eB\tanh(eBs)\,\eta^2\right]\right).
\end{eqnarray}
Performing the Matsubara sums, it can also be expressed as 
\begin{eqnarray}\label{nufinal}
    \nu(\eta)&=&-\frac{1}{4\pi^{3/2}}\int_0^\infty \frac{ds}{s^{5/2}}\frac{\mathrm{exp}\left[-eB\tanh(eBs)\eta^2\right]}{\sqrt{\tanh(eBs)/eBs}}\nonumber\\
		&&\times \frac{\partial}{\partial\mu}\Theta_3\left(\frac{1}{2}-\frac{i\mu}{2\pi T},\frac{i}{4\pi s T^2}\right).
\end{eqnarray}

The approximations for weak and strong limits of the magnetic field are obtained from Eq. (\ref{numero}).
For $|eB|\ll (\pi T)^2-\mu^2$,
\begin{multline}\label{nuweak}
    \nu(\eta) =  \frac{eB\eta}{\pi}T^2\bigg[\frac{|eB\eta|}{T}  \ln\!\left(\frac{1+e^{(|eB\eta|-\mu)/T}}{1+e^{\beta(|eB\eta|+\mu)/T}}\right)
   \\
   +\mathrm{Li}_2\!\left(-e^{(|eB\eta|-\mu)/T}\right)
    -\mathrm{Li}_2\!\left(-e^{(|eB\eta|+\mu)/T}\right)\bigg],
\end{multline}
where ${\rm Li}_2(x)$ is the dilogarithm function.

Notice that a very similar expression was derived in Ref.~\cite{Gorbar:2002iw} Eq.(74), replacing $eB\eta$ by a mass gap $\Delta$. This is totally expected since the relevant value to calculate the currents, as can be checked in Eq.(\ref{eq:nu}), is $\eta=y+y_\pm$, where $y_\pm=\frac{m_\pm}{eB}$. It means that the argument in the exponentials of the equation above is a function of $m_\pm$ only, which plays precisely the same role as $\Delta$: breaking sublattice symmetry.

For $|eB|\gg (\pi T)^2-\mu^2$,
\begin{equation}\label{nustrong}
    \nu(\eta)=\sqrt{|eB|}\,\frac{\mu}{\pi^{3/2}} e^{-|eB|\eta^2}.
\end{equation}
This is a remarkable result. 
Comparing with Eq. (\ref{jstrong}), we can see that $|j|=|e\nu|$.
In fact, integrating over the plane, we get
\begin{equation}\label{J-N5}
J_1 = |e|\text{sign}(B)N_5,
\end{equation}
where $J_1$ and $N_5$ are $j_1$ and $n_5$ integrated over the plane.
This is exactly the same result obtained in \cite{Fukushima:2008xe}, for the CME in QCD. 

\begin{figure}[t]
\centering
\includegraphics[width=0.4\textwidth]{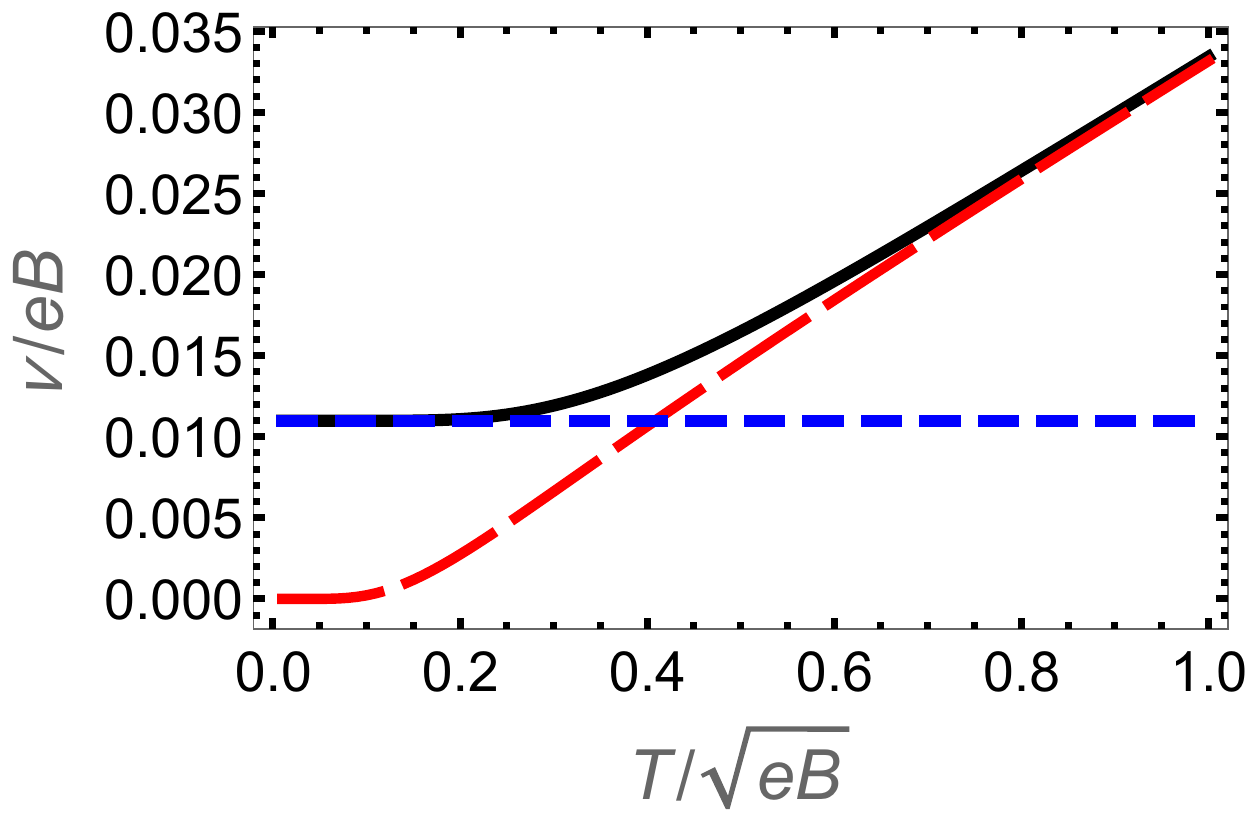}
\caption{
Behavior of $\nu(\eta)$ as a function of $T/\sqrt{eB}$ for fixed $\sqrt{|eB|}\eta = 0.5$ and $\mu/\sqrt{|eB|}=\pi/40$. 
The solid (black) curve corresponds to the numeric expression in Eq.~(\ref{nufinal}).
The long-dashed (red) curve corresponds to the weak field approximation in Eq.~(\ref{nuweak}).
The short-dashed (blue) curve to the strong field expansion in Eq.~(\ref{nustrong}). 
}
\label{fig3}
\end{figure}

The comparison between the strong and weak magnetic field limits and the full function is described in Fig.~\ref{fig3} for all parameters scaled with $|eB|$.

Similarly, the chiral separation effect is naturally obtained in our prescription. It is defined as a chiral charge separation that appears in the case a (non-chiral) chemical potential is present \cite{Metlitski:2005pr}. In the context of the quark gluon plasma it can be intuitively understood as follows: right-handed particles align with the magnetic field while their right-handed antiparticle anti-align with the field. If there is a chemical potential there will be a net current of right-handed particles in the direction of the field. In the same way, a current of left-handed particles is generated in the opposite direction, contributing with the same magnitude to the chiral current. Observing Eq.(\ref{nustrong}) and (\ref{jstrong}), we verify that:
\begin{equation}\label{J5-N}
J_5 = |e|\text{sign}(B)N.
\end{equation}

\subsection{Condensates}

Finally, we proceed with a similar reasoning to obtain condensates.
The non-vanishing condensates are 
\begin{align}
\sigma_3 =& \left \langle \bar{\psi}\gamma_3 \psi \right \rangle
            =-\text{tr}[\gamma_3 G(r,r)],\\
\sigma_{35} = &\left \langle \bar{\psi}\gamma_3\gamma_5 \psi \right \rangle
            = -\text{tr}[\gamma_3\gamma_5 G(r,r)],
\end{align}
generated through the gaps  $m_e$ and $m_o$. 
The reason to refer to these objects  as \emph{condensates} and not as a currents is because there is no dynamics along the $z$-direction. 
In this sense, $\sigma_3$ and $\sigma_{35}$ behave like order parameters for the symmetry breaking generated by the mass gaps in the sense that an expansion around these values naturally give a finite gap in the corresponding Lagrangian. 
It is convenient to write these condensates as
\begin{eqnarray}
    \sigma_3(y) &=& \sigma(y-y_+)+\sigma(y-y_-)\;,\\
    \sigma_{35}(y) &=& \sigma(y-y_+)-\sigma(y-y_-)\;,
\end{eqnarray}
where the function $\sigma$ in this case is
\begin{multline}\label{condensado}
    \sigma(\eta)=\frac{eB\,\eta}{\pi}
    T \sum_n \int_{0}^\infty \frac{ds}{s}
    \text{sech}^2(eBs)\left[\frac{eBs}{\tanh(eBs)}\right]^{1/2}
   \\
    \times \mathrm{exp}\left(-\left[s\left(\omega_n-i\mu\right)^2+eB\tanh(eBs)\eta^2\right]\right).
\end{multline}
We must be careful because the $\sigma$ function is UV divergent. 
Therefore we  separate the condensate in three contributions: the thermal-magnetic, magnetic, and vacuum part:
\begin{equation}
\sigma(\eta) = \sigma_{T,B}(\eta)+\sigma_B(y,m_\chi)+\sigma_0(m_\chi,\Lambda).
\end{equation}
We start by analyzing the thermal-magnetic contribution, which  is defined as
\begin{eqnarray}\label{sigmanum}
    \sigma_{T,B}(\eta)&=&\frac{eB\eta}{2\pi^{3/2}}\int_0^\infty \frac{ds}{s^{3/2}}\frac{\mathrm{exp}\left[-eB\tanh(eBs)\eta^2\right]}{\sqrt{\tanh(eBs)/eBs}}
		\nonumber\\
		&&\times\mathrm{sech}^2(eBs)\left[
		\Theta_3\left(\frac{1}{2}-\frac{i\mu}{2\pi T},\frac{i}{4\pi s T^2}\right)-1\right],
		\nonumber\\
\end{eqnarray}
where we have subtracted the $T,\mu=0$ part, as can be seen from the $\Theta_3-1$ term.

In the same way we proceeded with the currents and number densities, we can obtain approximations for weak and strong magnetic field from equation (\ref{condensado}), now taking care that we have to subtract the $T,\ \mu=0$ contribution.
For $eB\ll(\pi T)^2-\mu^2$,
\begin{multline}
    \sigma_{T,B}(\eta) =  -\frac{eB\eta}{\pi} T\bigg[\ln\left(1+e^{-\beta(|eB\,\eta|-\mu)}\right)\\
		+ \ln\left(1+e^{-\beta(|eB\,\eta|+\mu)}\right)
		\bigg].
\end{multline}
For $|eB|\gg(\pi T)^2-\mu^2$,
\begin{multline}
\sigma_{T,B}(\eta)= 
\frac{2eB\eta}{\pi}|2eB|^{1/4}T^{1/2}
e^{-|eB|\eta^2}\\
\times
\bigg[\text{Li}_{\frac{1}{2}}\left(-e^{-(\sqrt{2|eB|}-\mu)/T}\right)\\
+\text{Li}_{\frac{1}{2}}\left(-e^{-(\sqrt{2|eB|}+\mu)/T}\right)\bigg],
\end{multline}
where we also consider the fact that $T\lesssim \sqrt{|eB|}$, using the non-relativistic gas approximation. Here, ${\rm Li}_{\frac{1}{2}}(x)$ is the Poly-logarithmic function.

\begin{figure}[t]
\centering
\includegraphics[width=0.4\textwidth]{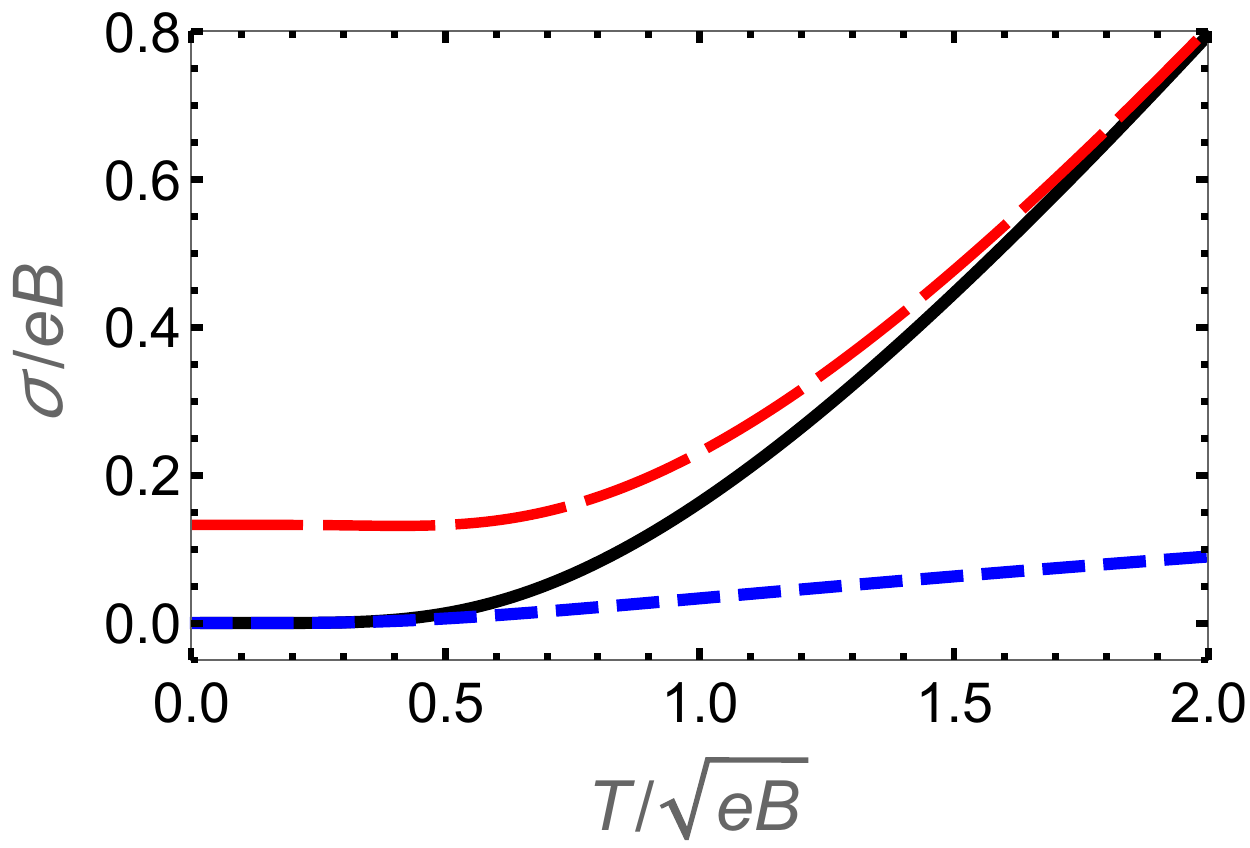}
\caption{
Behavior of $\tilde\sigma(\eta)$ as a function of $T/\sqrt{eB}$ for fixed $\sqrt{|eB|}\eta = 0.5$ and $\mu/\sqrt{|eB|}=\pi/40$.  
The solid (black) curve corresponds to the numeric expression in Eq.~(\ref{nufinal}).
The long-dashed (red) curve corresponds to the weak field limit in eq.~(\ref{nuweak}). 
The short-dashed (blue) curve to the strong field expansion in Eq.~(\ref{nustrong}). }
\label{fig4}
\end{figure}

The thermo-magnetic contribution to the condensates is described in Fig.~\ref{fig4}, compared with the weak and strong magnetic field approximations. 
The strong magnetic field (low temperature) approximation seems to be zero, but it is slowly growing with the temperature. 
This is not appreciable in this graph.

Now, the pure magnetic contribution is defined as 
\begin{eqnarray}
    \sigma_{B}(y,m_\chi)&=&\frac{1}{2\pi^{3/2}}\int_0^\infty \frac{ds}{s^{3/2}}\Bigg[
    \frac{\mathrm{exp}\left[-eB\tanh(eBs)\eta^2\right]}{\sqrt{\tanh(eBs)/eBs}}
    \nonumber\\&&
	\times eB\eta\,\text{sech}^2(eB s)-m_\chi e^{-m_\chi^2 s}\Bigg].
	\label{sigma_B}
\end{eqnarray}
where $\eta=y+m_\chi/eB$. The last term removes the vacuum part, so this function vanishes for $B=0$.

Until now, all the approximations have been made in terms of the magnetic field compared with the temperature. 
For the pure magnetic contribution, however, we have available two more energy scales to compare against, the masses (gaps) and the (inverse) of the length of the sample. 
Thus, to perform the proper-time integration and get a better comprehension about the scales for which the magnetic field strength can be considered weak or strong, we calculate the average of the condensate in space, defining it in the same way as in Eq. (\ref{Jpromedio}),
\begin{equation}\label{barsigma}
\bar\sigma_B(L_y,m_\chi) = \frac{1}{L_y}\int_{-L_y/2}^{L_y/2}dy\;\sigma_B(y,m_\chi).
\end{equation} 
Then, for $|eBL_y/2m_\chi|\ll 1$,
\begin{multline}
    \bar\sigma_B(L_y,m_\chi) = \text{sign}(m_\chi)\frac{(eB)^2}{12\pi m_\chi^2}\bigg[1+m_\chi^2L_y^2 \\
    +\left(\frac{eBL_y}{m_\chi^2}\right)^2 
\left(1+\frac{m_\chi^2 L_y^2}{2}\right)+\mathcal{O}\left(\frac{eBL_y}{2m_\chi}\right)^4\bigg],
\end{multline}
and for $|eBL/2m_\chi|\gg 1$,
\begin{equation}
    \bar\sigma_B(L_y,m_\chi) = \frac{|eB|L_y m_\chi}{\pi}\left[1+\frac{2|m_\chi|}{|eB|L_y}+\mathcal{O}\left(\frac{2|m_\chi|}{|eB|L_y}\right)^2\right].
\end{equation}
Finally, the third contribution is the vacuum part, which is divergent. Thus, 
 we introduce a UV cutoff $\Lambda$ to regulate the integrals, namely,
\begin{equation}
    \sigma_{0}(m_\chi,\Lambda)=\frac{m_\chi}{2\pi^{3/2}}\int_{\Lambda^{-2}}^\infty \frac{ds}{s^{3/2}}e^{-m_\chi^2 s}
    \approx \frac{m_\chi \Lambda}{\pi^{3/2}},
\end{equation}
assuming in the last step that $\Lambda\gg m_\chi$.

\subsection{Currents, densities and condensates at zero
temperature}

Although graphene is usually studied at room temperature, it is interesting to check the zero temperature limit. The calculation in this regime presents a technical issue whenever chemical potential and magnetic field are present in the Schwinger formalism: the Wick rotation in proper time is not trivial in this case, and therefore the propagator is highly oscillatory.
However, we can explore what happens at small magnetic field. 
In this case, in  Eq. (\ref{eq:propagator}) the trigonometric functions can be approximated as $\tan(eBs)\approx eB s$ and $\sec(eBs)\approx 1$.
\begin{align}
j(\eta) \to &\, \text{sign}(\mu)\frac{e^2B}{2\pi}\theta(|\mu|-|m_\chi|)\\
\nu(\eta) \to &\, \text{sign}(\mu)\frac{1}{2\pi} (\mu^2-m_\chi^2)\theta(|\mu|-|m_\chi|)\\
\sigma_\mu(\eta)\to& \frac{1}{4\pi}m_\chi(|\mu|-|m_\chi|)\,\theta(|\mu|-|m_\chi|),
\end{align}
where in the last equation we consider the chemical potential dependent term only and do not consider the vacuum part of the condensate.
It is interesting to see in this case that the magnetic conductivity is finite if only one pseudochirallity is present. 
For example if $|m_-|<\mu<|m_+|$, the electric current density is $j_1 = e^2B/2\pi$.

\section{Chiral chemical potential}
\label{sec:mu5}

In QCD, the CME is described by a single parameter, the so-called chiral chemical potential, which is proportional to the time derivative of the CS term. To establish our analogy, we have described a 2D Weyl-Dirac material  where the parity anomaly is manifested through the mass term $m_\pm$ in the Lagrangian. But because we need the system to be filled with charge carriers, the chemical potential also plays a role in the dynamics. If we consider the fermion system of our material degenerated with respect to the energy gaps $m_\pm$, we can explore what happens with the charge carriers, or holes, moving near the Fermi surface. There are two scenarios in this case, which we consider separately.

Let us turn-off the external magnetic field for a while. The Lagrangian~(\ref{lagrangiano})  can now be written as
\begin{equation}
{\cal L} = \sum_{\chi=\pm}\bar\psi_\chi[i\slashed{\partial}+\mu\gamma^0-m_\chi\gamma^3]\psi_\chi.
\end{equation}
From the equations of motion of each chiral field, we can obtain the dispersion relations for particles
\begin{equation}
p_0 = -\mu+\sqrt{\bm{p}^2+m_\chi^2}.
\label{FLdispersion}
\end{equation}
Thus, 
particles with opposite chiralities 
propagate differently. 
If the chemical potential is larger than the mass gaps, we can consider each specie propagating near their Fermi surface. 
In this case, \emph{antiparticles} can be omitted in the discussion because those states are already filled by assumption. 
Now, the Fermi momentum for each chirality is:
\begin{equation}
p_\chi = \sqrt{\mu^2-m_\chi^2}.
\end{equation}
Expanding around it, we get
\begin{equation}
p_0\approx  \pm v_\chi |\bm{p}-\bm{p}_\chi|\;,
\end{equation}
where now $+$ and $-$ label quasi-particles and their anti-particles (quasi-holes), respectively. We have restored the velocity units to make clear the dynamics.
Then, we observe that we must define a new Fermi velocity for each chirality, namely,
\begin{equation}
v_\chi = \sqrt{1-m_\chi^2/\mu^2}.
\end{equation}
Thus, quasi- particles and holes propagate at different velocities under this perspective.

We can see that the Lagrangian that describes the motion of our Fermi liquid system reads
\begin{equation}
{\cal L} = \sum_{\chi=\pm}\bar\psi'_\chi[i\gamma^0\partial_0-v_\chi\bm{\gamma}\cdot\bm{\nabla}]\psi'_\chi\;,
\end{equation}
which describes a  gapless fermion system, with quasi- particles and holes moving  differently. Moreover,  the  corresponding fields in momentum space are also different,  
\begin{equation}
\psi'_\chi(p_0,\bm{p}) = \psi_\chi(p_0,\bm{p}+\bm{p}_\chi).
\label{newpsi}
\end{equation}
This can be seen as if the Dirac points described in Fig.~\ref{red} were displaced as $\bm{K}\to \bm{K}+\bm{p}_+$ and $\bm{K}'\to \bm{K}'+\bm{p}_-$.

\begin{figure}
\includegraphics[scale=.26]{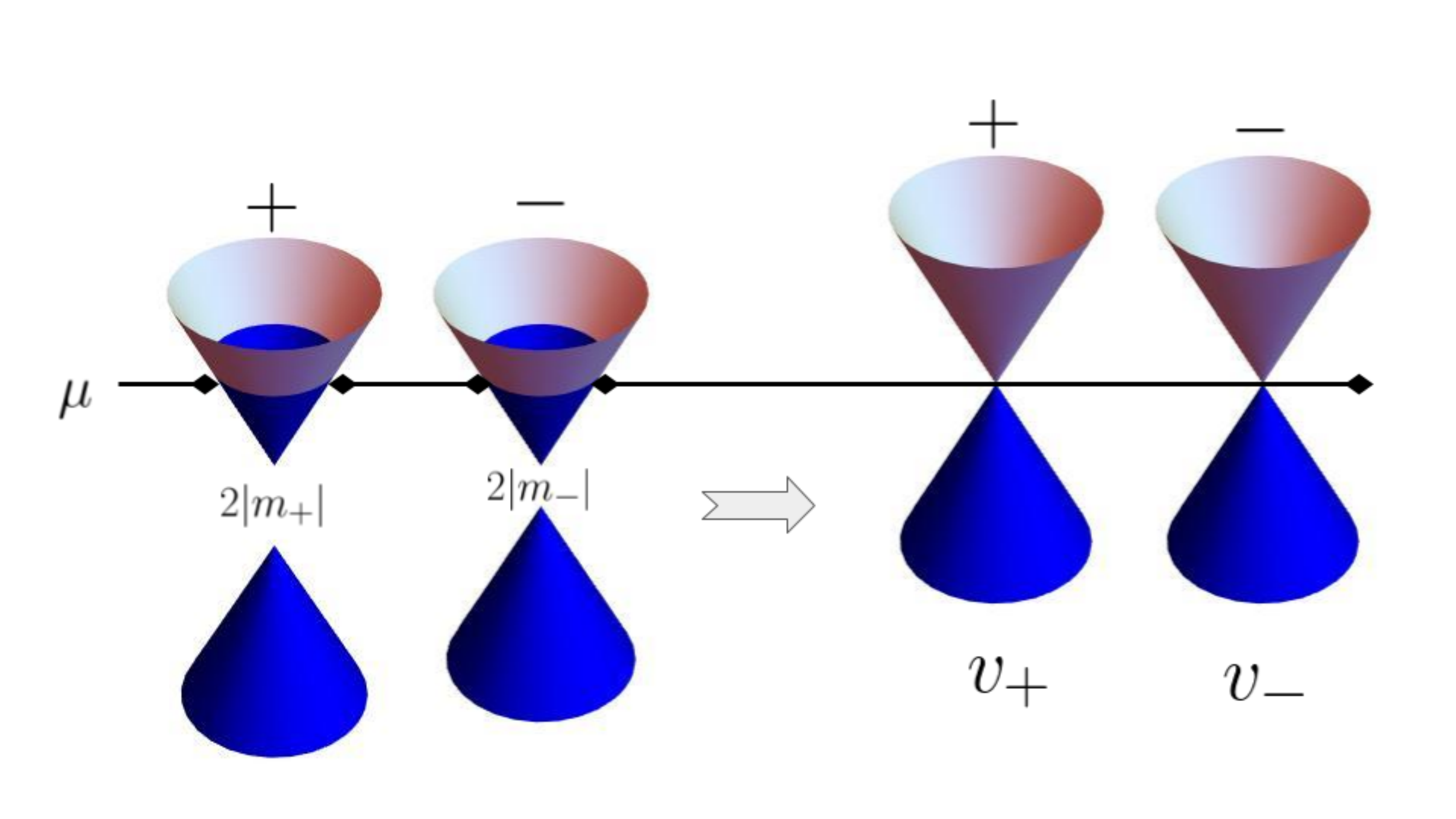}
\caption{Schematic description of the new Dirac cones considering charge carriers propagating near the the Fermi surface, considering with different Fermi velocity for each pseudochirality}
\label{conosA}
\end{figure}

\bigskip
Let us explore this approximation in another way, now by assuming that all the quasi- particles and holes propagate with the same Fermi velocity  $v_F'$. Thus, the Fermi momentum for each chirality can be defined as
\begin{equation}
p_\chi = \frac{m_\chi v_F'}{\sqrt{1-{v_F'}^2}}\;.
\end{equation}
Expanding Eq.~(\ref{FLdispersion}) around $p_\chi$ we obtain
\begin{equation}
p_0\approx -\mu+\frac{|m_\chi|}{\sqrt{1-{v_F'}^2}}\pm v_F' |\bm{p}-\bm{p}_\chi|\;.
\end{equation}
In this way, we obtain the above dispersion relations from the effective Lagrangian
\begin{equation}
{\cal L} =\bar\psi'[i\gamma^0\partial_0
-v_F'\bm{\gamma}\cdot\bm{\nabla}
+\mu'\gamma^0+\mu_5\gamma^0\gamma^5]\psi',
\end{equation}
with the new fields defined in Eq. (\ref{newpsi}), and where the chemical potentials are defined as 
\begin{align}\label{mu5}
\mu' &=  \mu -\frac{|m_-|+|m_+|}{2\sqrt{1-{v_F'}^2}},\\
\mu_5 &=  \frac{|m_-|-|m_+|}{2\sqrt{1-{v_F'}^2}}.
\end{align}
At this moment, we have not specified the orientation of the Fermi momentum because the Fermi velocity is unknown. 
Certainly, the velocity of propagation of the charge carriers we consider is near the original average Fermi energy $\mu$.
Therefore, we can set $\mu'=0$ and what we get is an explicit relation of the Fermi velocity and chiral chemical potential,
\begin{align}
v_F'&=\sqrt{1-\left(\frac{|m_-|+|m_+|}{2\mu}\right)^2}, \\
\mu_5 & = \mu\frac{|m_-|-|m_+|}{|m_-|+|m_+|}.
\label{eq:mu5}
\end{align}
From the above expressions, we observe that mass gaps  can be parametrized in terms of this single chiral chemical potential, just as in the CME. 

\begin{figure}
\includegraphics[scale=.26]{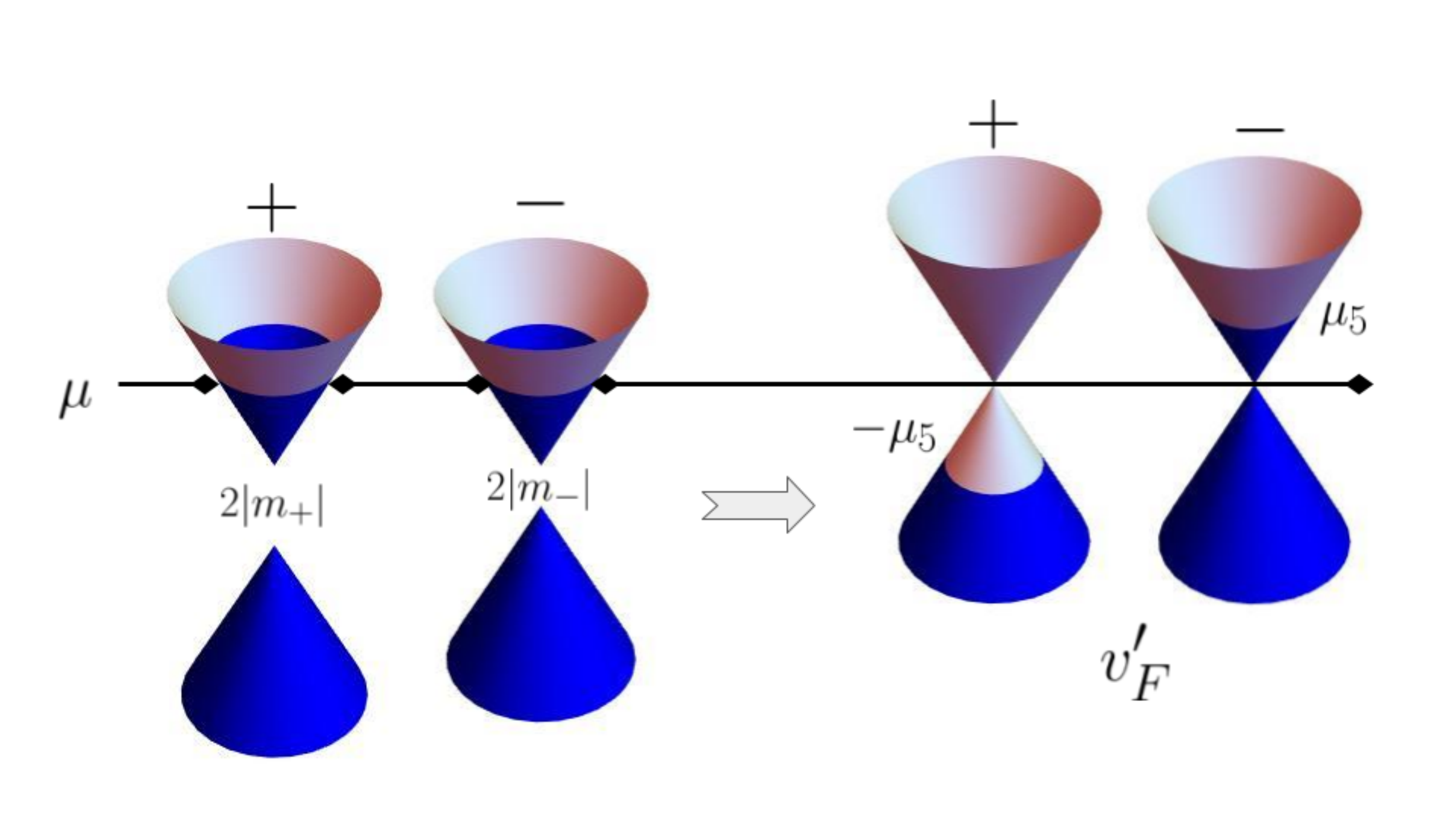}
\caption{Schematic description of the new Dirac cones considering charge carriers propagating near the the Fermi surface, considering a new average Fermi velocity for both pseudochiralities.}
\label{conosB}
\end{figure}

This effective Lagrangian corresponds to remove the masses and add a chemical potential $\pm \mu_5$ to each chirality.).
In particular, in the case of zero temperature and small magnetic field, we have
\begin{align}
j_1\to &\, \text{sign}(\mu_5)e^2B/\pi\\
n_5\to &\, \text{sign}(\mu_5)\mu_5^2/\pi
\end{align}
and $n,\,j_{15},\,\sigma_3,\,\sigma_{35}\to 0$.
This cancellation occurs because there is the same amount of fermion with one pseudo-chirality and anti-fermions with opposite pseudo-chirality.

Let us emphasize once more that our construction is based on the symmetry breaking of Dirac Points which in those planar systems with relativistic fermions play the role of pseudo-chirality. As such, $\mu_5$ encodes this asymmetry rather than being related to actual chirality.

\section{Discussion and conclusions} \label{conclusions}

In this article we have established an analogy of the CME for 2D Dirac-Weyl matter. We have considered a system of fermions with dynamics restricted to a plane which have two different mass gaps. 
One of those mass gaps is known to be related with the CS term in QED$_3$. It is under investigation if the same holds for PQED or RQED.
The other mass gap $m_e$ breaks the equivalence between the two  triangular sublattices of the underlying honeycomb array. 
When an external magnetic field is applied parallel to this system, the generation of an electric current along the direction of the magnetic field is formed. This corresponds to an analogue of the CME, but in terms of the pseudo-spin of the charge carriers of our system. 
We have further explored the formation of number density, chiral number density, axial current and condensates associated to the mass gaps.

The most remarkable relation and close connection of the CME and the effect discussed in here is represented in Eq.~(\ref{J-N5}), which was obtained in the strong magnetic field (low temperature) regime and is functionally the same obtained for the ordinary CME in QCD~\cite{Fukushima:2008xe}. Let us stress that in the latter case, the phenomenon is believed to occur at high-temperature.
The CME is described through a chiral chemical potential, while in our case we need two mass gaps. 
This can be understood considering that in graphene-like system first we need a mechanism that distinguishes the left and right pseudo-chiralities and further another one that breaks the balance.

Considering our Fermi system to be degenerate, it is possible to describe it as a Fermi liquid effective model, and the chiral imbalance can be seen in two ways: as quasi-particles with different pseudo-chiralities propagating at different Fermi velocities, or like a fluid whose imbalance is produced by a chiral chemical potential that depends on the difference of the mass gaps, as can be seen from Eq. (\ref{mu5}).
  In both cases, the consequence is a shift in the Dirac points, hinting a deformation of the crystal structure. 
  This observation reinforces the fact that strained graphene is a good candidate to reproduce this effect.
  
  It is interesting to remark  that different arrangements of a graphene membrane can mimic many QCD situations. 
  We want to explore in more detail the mass gap generation, including the possibility of generating it through external fields, which requires a proper treatment of the gauge sector.
  This idea is under scrutiny and results shall be presented elsewhere.

\section*{Acknowledgements}
This work was supported by FONDECYT (Chile) under
grant numbers 1150847, 1130056, 1150471 and 1170107.
AR is supported by Consejo Nacional de Ciencia y Tecnolog\'ia (Mexico) Grant 256494.
AJM is supported by BELSPO-Belgium and FAPESP under grant 2016/12705-7. 
We acknowledge the research group GI\,172309\,C \emph{Cosmolog\'ia y Part\'iculas Elementales} at UBB and G. Krein at IFT-UNESP for the kind hospitality. AJM ackowledges D.~Dudal for fruitful discussions.

\appendix

\section{Some formulas}\label{formulas}

The Jacobi $\Theta_3$ function used for currents masses and condensates is defined as 
\begin{equation}
\Theta_3(z,\tau) = \sum_{n=-\infty}^{\infty} e^{i\pi(\tau n^2+2z n)}.
\end{equation}
This inversion formula is
\begin{equation}
\Theta_3(z,\tau) =\frac{e^{-i\pi z^2/\tau}}{\sqrt{-i\tau}}
\Theta_3\left(\frac{z}{\tau},\frac{-1}{\tau}\right).
\end{equation}

\bigskip
Some results are expressed in terms the polylogarithm function, defined as
\begin{equation}
\text{Li}_s(z) = \sum_{n=1}^\infty \frac{z^n}{n^s}.
\end{equation}

\bigskip

A useful trick used in this article is to include a \emph{momentum} integral in order to remove some denominator in the Schwinger proper time, which allow us to integrate the proper analytically. 
For this reason, in some approximations, the proper-time denominator is exponentiated using the formula
\begin{equation}\label{id}
\frac{1}{s^{n/2}} = \frac{2}{\Gamma\left(\frac{n}{2}\right)}\int_0^\infty dp\ p^{n-1} e^{-sp^2}.
\end{equation}

\section{Weak Field Approximation}

For the weak field approximation of $j(\eta)$, $\nu(\eta)$ and $\sigma_{T,B}(\eta)$, we first start from Eq. (\ref{corriente}), (\ref{numero}) and (\ref{condensado}), respectively.
First, we notice that for $eB\ll (\pi T)^2 - \mu^2$, we can set ${\rm sech}(eBs)\simeq 1$ and $\tanh{(eBs)}\simeq eBs$, allowing the exact integration in proper-time in the case of $j_1(\eta)$. 
For $\nu(\eta)$ and $\sigma(\eta)$ there is a $s^{-1}$ factor that can 
replaced by the integral in a momentum variable described in Eq. (\ref{id}) with $n=2$.
After integration in proper time, the Matsubara frequencies can be summed exactly and then we perform the integral in terms of the auxiliary $p$ in the case of number densities and condensates.

In the case of the pure magnetic contribution to the condensates, we average first on the space as is indicated in Eq.(\ref{barsigma}). 
Then we expand in powers of $eB$ and finally integrate on the proper time. 
The expansion itself shows the scale of the expansion, which is $m_\chi^2/L_y$.

\section{Strong Field Approximation}

For the strong field approximation of $j(\eta)$, $\nu(\eta)$ and $\sigma_{T,B}(\eta)$, we start as in the previous case from Eq. (\ref{corriente}), (\ref{numero}) and (\ref{condensado}), respectively.
Now, we consider the case of extremely intense field, $eB\gg (\pi T)^2 - \mu^2$.  
In this regime,  we notice that $\tanh(|eB|s)\approx 1$, 
and in the case of the condensate, sech$(eBs)\approx 4e^{-2|eB|}$. 
This generates an denominator $s^{-1/2}$ in the three cases, and we replace it by an integral in momentum described in Eq. (\ref{id}) with $n=1$. 
Now, the integral in proper-time can be solved exactly and also the sum in Matsubara frequencies. 
Then we integrate the auxiliary momentum term.

In the case of the thermomagnetic contribution for the condensates, the non-relativistic approximation was used. 
This can be applied when the mass is larger or of the same order than the temperature. 
Here, the mass term comes from the coth term and is $\sqrt{2|eB|}$. 
Then, with the condition $T\lesssim |2eB|^{1/2}$ is enough to make this approximation.

For the case of the pure magnetic contribution to the condensates, the procedure is a bit more complicate. 
We first change the proper-time integration variable time with one scaled the magnetic field, $\bar s = |eB| s$. 
Then, we separate the integral, one from $0<\bar s <1$ and the other with $1<\bar s <\infty$. 
In the small $\bar s$ integral, we can approximate $\tanh(\bar s)\approx \bar s$ and sech$^2(\bar s)\approx 1$.
In the high $\bar s$ integral, we can set tahh$(\bar s)\approx 1$ and sech$^2(\bar s)\approx 4e^{-2\bar s}$.
Now, the integrals in proper-time can be readily done, with the tricks of the momentum integral in Eq. (\ref{id}), average with respect to the space as Eq. (\ref{barsigma}) and expand in terms of $|eB|^{-1}$.

%

\end{document}